\documentclass{mem}
\usepackage{natbib}\usepackage{txfonts}\usepackage{balance}
\usepackage{graphicx}
\usepackage{txfonts}
\usepackage[a4paper]{hyperref}
\idline{79}{3}
\begin{document}
\def\teff{$T\rm_{eff }$}
\def\kms{$\mathrm {km s}^{-1}$}

\title{
Open Clusters in the log Age vs. M$_V$ plane.
}


\author{
M. \,Bellazzini\inst{1}, S. \,Perina\inst{1,2}, S. \,Galleti\inst{1},
L. \,Federici\inst{1}, A. \,Buzzoni 
\and F. \,Fusi Pecci}

  \offprints{M. Bellazzini}

\institute{
Istituto Nazionale di Astrofisica --
Osservatorio Astronomico di Bologna, Via Ranzani 1,
I-40127 Bologna, Italy
\email{michele.bellazzini@oabo.inaf.it}
}

\authorrunning{Bellazzini et al.}

\titlerunning{Open Clusters in the log Age vs. M$_V$ plane}

\abstract{In the log Age vs. integrated absolute magnitude ($M_V$) 
plane, the open clusters
of the Milky Way form a well-defined band parallel to theoretical sequences 
decribing the passive evolution of Simple Stellar Populations and display a
pretty sharp upper threshold in mass ($M\sim 2\times 10^4 ~M_{\odot}$) over a
4 dex range of ages.  
}
\maketitle{}

\section{Introduction}

The evolution of integrated spectro-photometric properties of a Simple Stellar
Population \cite[SSP, i.e. an idealized population of stars having the same chemical
composition and the same age,][]{rfp} is one key prediction of stellar
theoretical models \citep[see, for example][and references therein]{buz,cla1}.
In particular, it is well known that the total luminosity of a 
SSP must decrease with time as massive stars 
progressively exhaust their nuclear fuel and conclude their evolutionary 
lifetime, thus ceasing to contribute to the luminosity of the SSP.

\begin{figure}[t!]
\resizebox{\hsize}{!}{\includegraphics[clip=true]{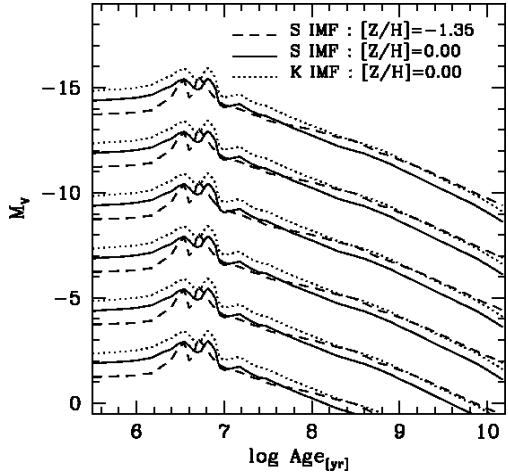}}
\caption{\footnotesize
Passive evolutionary sequences for SSPs of different metallicities ([Z/H]) and
IMFs \cite[S = Salpeter; K = Kroupa, see][]{krou}, from \citet{cla1,cla2}.
Each bundle of three sequences correspond to a given total mass.
}
\label{teo}
\end{figure}

In Fig.~\ref{teo} we show various theoretical evolutionary sequences describing
the fading with age of SPSSs \cite[from][]{cla1,cla2}, in the plane of the
logarithm of the SSP age versus its integrated absolute V magnitude ($M_V$),
hereafter A-M$_V$ diagram, for brevity 
\cite[see][and references therein, for the application of this or similar
diagrams to the study of star clusters in different environments]{giel,whit}.
It can be appreciated that (i) for ages $>10^7$ yr the evolutionary sequences
are essentially linear ($M_V\propto 1.8\times {\rm log} Age_{[yr]}$, with $a\simeq
1.8$), and, (ii) the sequences depends quite weakly on the
assumed metallicity and/or Initial Mass Function (IMF) of the SSP.
Once a metallicity and a form of the IMF are assumed, each sequence directly
correspond to a total stellar mass; thus the mass of {\em real} SSPs can be
compared in this plane independently of their respective age. Moreover, the
past and future evolution of such SSPs can be directly read on this diagram.
Given the weak dependence on age and IMF, in the following we will adopt a grid
of solar metallicity / Salpeter-IMF sequences. These define a 
total-stellar-mass scale whose zero point may be uncertain up to a factor of 
a few, while {\em mass differences} should be pretty reliable and 
{\em homogeneous}.   
Star clusters are the best approximation of SSPs available in nature.
Classical Globular Clusters (GC) are all very old and should lie in a narrow
slice of the A-M$_V$ diagram. Here, for simplicity, we adopt Age = 12 Gyr 
\citep{raf} for all the Galactic GCs, for which we took $M_V$ from
\citet{harris}. On the other hand,  Galactic Open Clusters (OC) are known to
span a large range in ages (from millions to billions years). For their sparse
nature, it is quite hard to obtain reliable integrated properties of OCs;
nevertheless the WEBDA database\footnote{\tt www.univie.ac.at/webda/} 
collects also
OC $M_V$ from many different sources and, in general, the agreement between
independent estimates is reassuringly good. We extracted, from WEBDA, ages and
$M_V$ for 293 OCs, taking the $M_V$ estimates from \citet{lata}, \citet{bat94},
\citet{spas}, \citet{pand}, and \citet{sagar}, in order of preference.  

\begin{figure*}[]
\resizebox{\hsize}{!}{\includegraphics[clip=true]{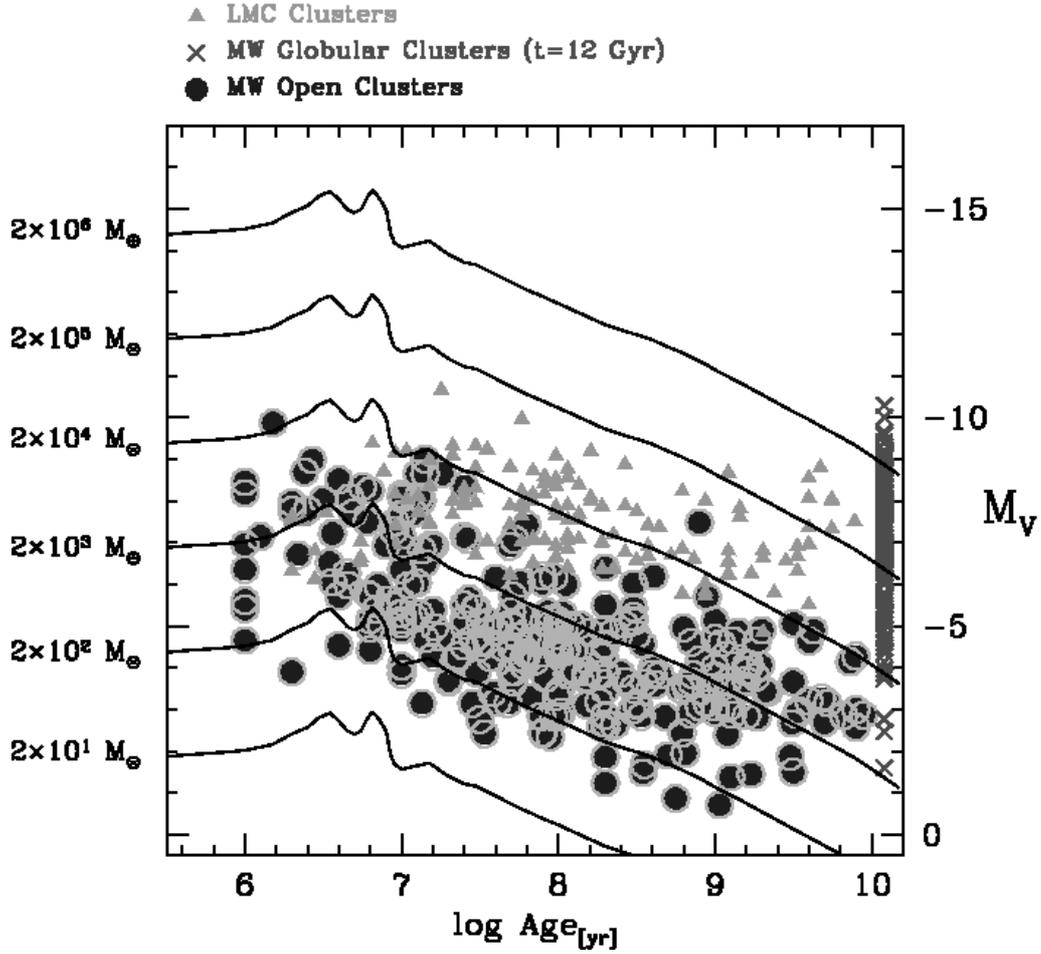}}
\caption{\footnotesize
Galactic GCs and OCs and LMC clusters in the A-$M_V$ plane. The
passive-evolution sequences are for solar metallicity and Salpeter's IMF
\cite[from][]{cla1,cla2}. The only OC clearly exceeding the $2\times10^4
~M_{\odot}$ threshold is Tombaugh~2, around log Age $\sim 9$.}
\label{lmc}
\end{figure*}
\begin{figure}[t!]
\resizebox{\hsize}{!}{\includegraphics[clip=true]{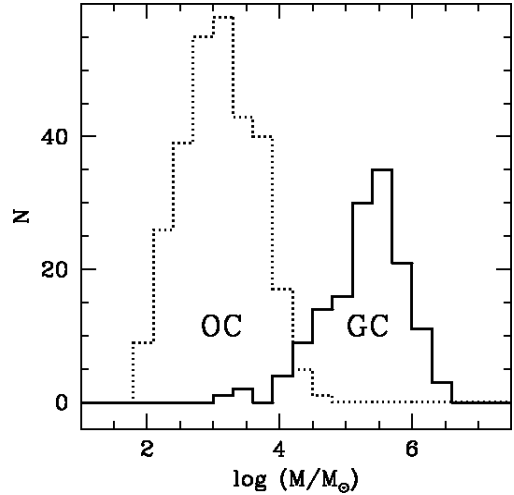}}
\caption{\footnotesize
Mass distribution of Galactic OCs and GCs, from
interpolation on the theoretical grid of Fig.~\ref{lmc}.
}
\label{masse}
\end{figure}

In Fig.~\ref{lmc} Galactic OCs are compared to GCs and to stars cluster of the
Large Magellanic Cloud (data from \citet{syd}, treated as in
\citet{ffp05}), in the A-M$_V$ diagram. 
It is interesting to note that OCs form a well defined band, parallel to the
evolutionary sequences and approximately comprised between 
$M\simeq 5\times 10^1 ~M_{\odot}$ and $M\simeq 2\times 10^4~M_{\odot}$. 
The different distribution of LMC clusters
demonstrate that the occurrence of a mass threshold is not universal, but 
it is likely associated with the particular environment in which clusters 
formed. A thorough discussion of the mechanisms that shape the distribution of
cluster populations in this plane can be found in 
\citet[][see also references therein]{whit}.

Fig.~\ref{lmc} also recalls that OCs and GCs have two well separated
mass distributions; while the difference in mean mass is
obviously not a surprise, the bimodality of the mass distribution of Galactic
star clusters as a whole (OC+GC) is far from trivial 
\cite[see Fig.~\ref{masse}, and][]{syd2}. Finally, it is interesting to note that, 
at the dawn of the Galactic era, the progenitors of GCs had
luminosities typical of dwarf galaxies ($-10\le M_V\le-15$, approximately).

\begin{acknowledgements}
M.B acknowledges the financial support to this research by INAF, through the
grant CRA 1.06.08.02.
\end{acknowledgements}

\bibliographystyle{aa}

\end{document}